\documentclass[prb]{revtex4}

\usepackage[dvips]{graphicx}
\usepackage{dcolumn}
\usepackage{bm}
\begin{document}


\title{Optical properties of gold films and the Casimir force}

\author{V. B. Svetovoy}
\email[Corresponding author:]{v.b.svetovoy@ewi.utwente.nl}
\affiliation{MESA+ Research Institute, University of Twente, PO 217,
7500 AE Enschede, the Netherlands}
\author{P. J. van Zwol}
\affiliation{Department of Applied Physics, Netherlands Institute
for Metals Research and Zernike Institute for Advanced Materials,
University of Groningen, Nijenborgh 4, 9747 AG Groningen, the
Netherlands}
\author{G. Palasantzas}
\affiliation{Department of Applied Physics, Netherlands Institute
for Metals Research and Zernike Institute for Advanced Materials,
University of Groningen, Nijenborgh 4, 9747 AG Groningen, the
Netherlands}
\author{J. Th. M. De Hosson}
\affiliation{Department of Applied Physics, Netherlands Institute
for Metals Research and Zernike Institute for Advanced Materials,
University of Groningen, Nijenborgh 4, 9747 AG Groningen, the
Netherlands}


\begin{abstract}
Precise optical properties of metals are very important for accurate
prediction of the Casimir force acting between two metallic plates.
Therefore we measured ellipsometrically the optical responses of
$Au$ films in a wide range of wavelengths from 0.14 $\mu m$ to 33
$\mu m$. The films at various thickness were deposited at different
conditions on silicon or mica substrates. Considerable variation of
the frequency dependent dielectric function from sample to sample
was found. Detailed analysis of the dielectric functions was
performed to check the Kramers-Kronig consistency, and extract the
Drude parameters of the films. It was found that the plasma
frequency varies in the range from 6.8~$eV$ to 8.4~$eV$. It is
suggested that this variation is related with the film density.
X-ray reflectivity measurements support qualitatively this
conclusion. The Casimir force is evaluated for the dielectric
functions corresponding to our samples, and for that typically used
in the precise prediction of the force. The force for our films was
found to be 5-14\% smaller at a distance of 100~$nm$ between the
plates. Noise in the optical data is responsible for the force
variation within 1\%. It is concluded that prediction of the Casimir
force between metals with a precision better than 10\% must be based
on the material optical response measured from visible to
mid-infrared range.
\end{abstract}

\maketitle

\section{\label{Sec1}Introduction}
The force between electrically neutral metallic plates separated by
a small vacuum gap (of micron/submicron dimensions), as predicted by
the eminent Dutch physicist H.B. Casimir in 1948, still attracts
considerable interest. On one hand this interest is curiosity driven
since the force is connected with the zero-point fluctuations in
vacuum, and on the other hand the interest stems from practical
applications because modern microtechnology approaches the limit
where the force starts to influence the performance of microdevices.
During the past decade the Casimir force was measured with
increasing precision in a number of experiments using different
techniques such as torsion pendulum \cite{Lam97}, atomic force
microscope (AFM) \cite {Moh98,Har00}, microelectromechanical systems
(MEMS) \cite {Cha01,Dec03a,Dec03b,Dec05} and different geometrical
configurations: sphere-plate \cite{Lam97,Har00,Dec03b}, plate-plate
\cite{Bres02}, crossed cylinders \cite {Ede00}. In most cases the
bodies were covered with gold evaporated or sputter deposited to a
thickness of 100-200 $nm$.

Relatively low precision, 15\%, in the force measurement was reached
for the plate-plate configuration \cite{Bres02} because of the
parallelism problem. In the torsion pendulum experiment \cite
{Lam97} the force was measured with an accuracy of 5\%. In the
experiments \cite{Har00,Ede00,Cha01} errors were claimed on the
level of 1\%. In the most precise up to date experiment
\cite{Dec03a,Dec03b,Dec05} the experimental errors claimed to be as
low as 0.5\%.

Comparison between theory and experiment answers an important
question: how accurately do we understand the origin of the force?
To make a precise evaluation of the force taking into account real
conditions of the experiments is equally difficult as to make a
precise measurement. In its original form, the Casimir force
\cite{Cas48} given by
\begin{equation}
F_{c}\left( a\right) =-\frac{\pi ^{2}}{240}\frac{\hbar c}{a^{4}}
\label{Fc}
\end{equation}
was calculated between the ideal metals at zero temperature. It
depends only on the fundamental constants and the distance between
the plates $a$. The force between real materials was derived for the
first time by Lifshitz \cite {Lif56,DLP,LP9}. The material
properties enter the Lifshitz formula via the frequency dependent
dielectric function $\varepsilon \left( \omega \right) $. This
formula became the basis for all precise calculations of the force.
Corrections to Eq.~(\ref{Fc}) can be very large especially at small
separations ($<100\;nm$) between bodies.  The Lifshitz formula
accounts for real optical properties of the materials, and for
finite temperature effects. An additional source of corrections to
the force is the surface roughness of interacting plates
\cite{Kli96,Gen03,Mia05}.

In all the experiments mentioned above the bodies were covered with
metallic films but the optical properties of these films have never
been measured. It is commonly accepted
\cite{Har00,Dec03b,Dec05,Kli00} that these properties can be taken
from the handbooks tabulated data \cite{HB1,Wea81} together with the
Drude parameters, which are necessary to extrapolate the data to low
frequencies. This might still be a possible way to estimate the
force, but it is unacceptable  for calculations with controlled
precision . Lamoreaux \cite{Lam99b} was the first who recognized
this problem. The reason is very simple
\cite{Lam99b,Lam99a,Sve00a,Sve00b}: optical properties of deposited
films depend on the method of preparation, and can differ
substantially from sample to sample.

Recently analysis of existing optical data for $Au$ was undertaken
\cite{Pir06} to explore how significant is the effect of variation
of the optical properties on the Casimir force. It was demonstrated
that different sets of the data deviate considerably. This variation
influences the Casimir force on the level of 5\% in the distance
range of the most precise experiments. Significant sample dependence
of the force raises doubts on reported agreement between theory and
experiment within 1\% precision \cite{Har00,Dec05}. This is an
important issue that has to be further thoroughly investigated.

In this paper we present the optical properties of $Au$ films of
different thickness deposited in two different evaporators on
silicon and/or mica substrates, unannealed or annealed after the
deposition. Moreover, we will discuss the influence of measured
optical properties of gold films on the precise evaluation of the
Casimir force. For the first time the characterization of the films
was performed over a wide frequency range. It was done
ellipsometrically using the infrared variable angle spectroscopic
ellipsometer in the wavelength range 1.9 - 33~$\mu m$, and the
vacuum ultraviolet ellipsometer in the range 0.14 - 1.7 $\mu m$. In
addition, the film roughness was characterized with atomic force
microscopy (AFM), and the electron density in the films was
estimated from X-ray reflectivity measurements.

Careful analysis of the data is performed to extract the values of
the Drude parameters. It includes joint fits of the real and
imaginary parts of the dielectric function, or refractive index and
extinction coefficient in the low frequency range; the
Kramers-Kronig consistency of the dielectric function or complex
refractive index performed at all frequencies. The most important
conclusions that follow from this analysis is that the films
deposited at the same conditions, but having different thicknesses,
have considerably different dielectric functions; annealing or
change of the deposition method showed also influence on the optical
properties. At any rate this difference cannot be ignored in a
precise calculation of the Casimir force. We demonstrate that the
optical data typically used for the force evaluation in former
studies are far away from that found in our samples. The main reason
for this deviation is the use of the Drude parameters, which
correspond to a perfect gold single crystal rather than real
polycrystalline films, containing a number of different defects.

The paper is organized as follows. In Sec. \ref{Sec2} we describe
preparation and characterizations of $Au$ films and make comparison
with results known from literature. Analysis of the optical data is
presented in Sec. \ref{Sec3}, where the Drude parameters, and
uncertainties in these parameters are determined. Calculation of the
Casimir force for our samples is given in Sec. \ref{Sec4}. Our
conclusions are presented in the last Section.

\section{\label{Sec2}Experimental}
Five gold films were prepared by $Au$ deposition on cleaned (100)
$Si$ substrates and freshly cleaved mica. The native oxide on $Si$
substrates was not removed; the root-mean-square (rms) roughness was
0.3 $nm$ for the $Si$ substrates, while the mica substrate was
atomically flat. The first three samples (numbered as 1, 2, and 3)
were prepared on $Si$ covered first with 10 $nm$ adhesive sublayer
of titanium followed by deposition of 400, 200, and 100 $nm$ of $Au$
from the source of 99.999\% purity. The electron-beam evaporator was
used for deposition at a base pressure of $10^{-6}\;mbar$. The
deposition rate was 0.6 $nm/s$. The temperature of the samples was
not controlled in the evaporator and it was approximately at room
temperature. The other two samples were prepared in a thermal
evaporator at the same base pressure and deposition rate. One film
was deposited to a thickness of 120 $nm$ on Si with chromium
sublayer (sample 4). The other film of the same thickness was
deposited on mica, annealed at $375\;^{\circ}C$ (2 hours) and slowly
cooled down in a period of 6 hours resulting in an atomically flat
film (sample 5).

\begin{figure}[tbp]
\includegraphics[width=17cm]{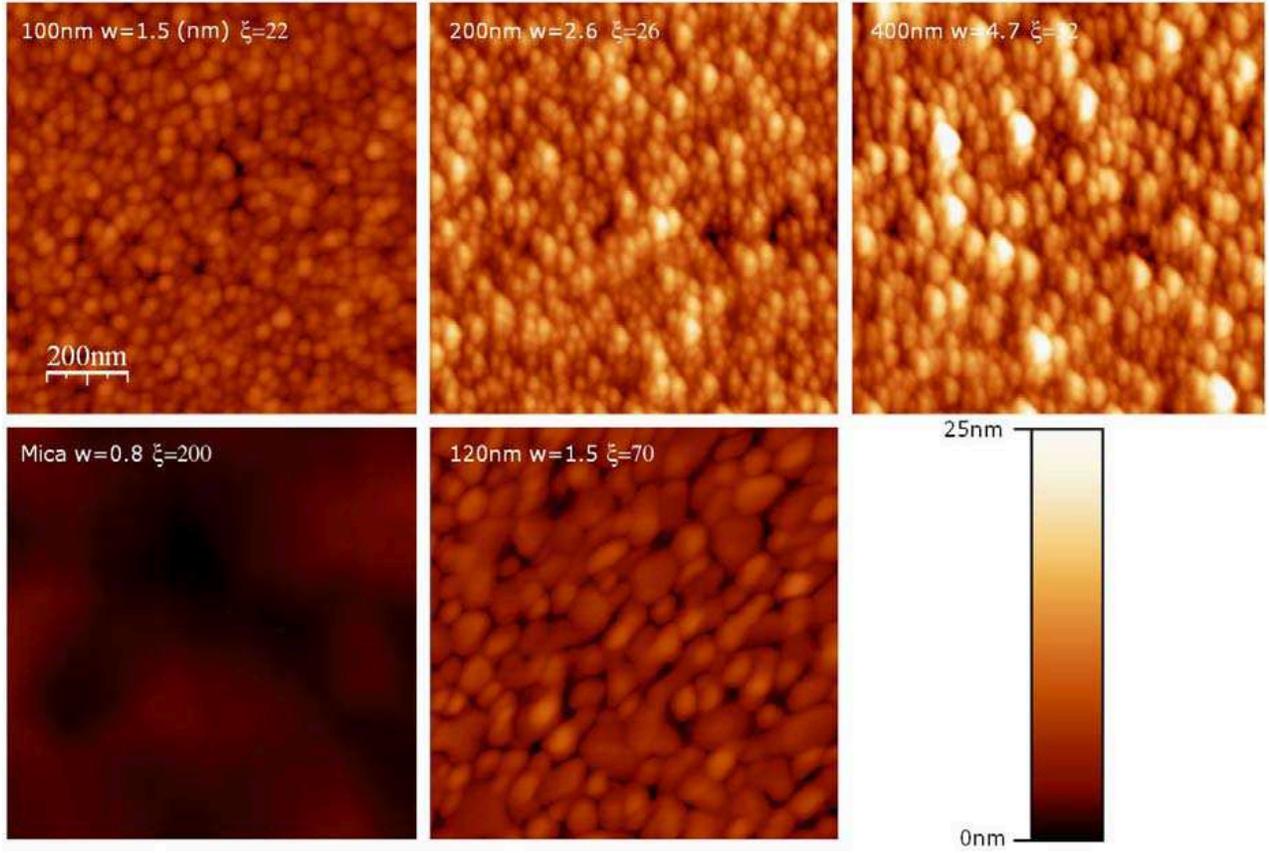}\newline
\caption{(Color online) Surface scans of all films with AFM using
the same color scale. The scan area is $1\;\mu m^2$. The film
thickness, rms roughness $w$, and correlation length $\xi$ are shown
in each panel (in $nm$).} \label{fig1}
\end{figure}

The Atomic Force Microscope  (Veeco Dimension 3100) was used to
determine the surface morphology. The roughness scans are shown in
Fig.~\ref{fig1} for all 5 samples. The corresponding rms roughness,
$w$, and correlation length \cite{Mea94,Kri95,Zho01} $\xi$, shown in
each panel, were obtained as the average values found from multiple
scans. The correlation lengths for the first 3 samples corresponding
to the lateral feature sizes were reported before \cite{Zwo07}. It
should be noted that for sample 4 (120 $nm$ $Au/Si$) the correlation
length is larger than those for the other three films on $Si$. It
can result from differences in the evaporation process or due to
different adhesive layer. The annealed film on mica had very smooth
hills and the largest correlation length.

\bigskip

Optical characterization of the films was performed by J. A. Woollam
Co., Inc. \cite{Woo}. The vacuum ultraviolet variable angle
spectroscopic ellipsometer (VUV-VASE) was used in the spectral range
from 137 $nm$ to 1698 $nm$ at two angles of incidence $65^{\circ}$
and $75^{\circ}$ ($\pm 0.01^{\circ}$). The steps in the wavelength
$\lambda$ was increased quadratically with $\lambda$ from 1.5 $nm$
to 200 $nm$. In the spectral range from 1.9 $\mu m$ to 32.8 $\mu m$
the infrared variable angle spectroscopic ellipsometer (IR-VASE) was
used at the same incidence angles in steps $\sim\lambda^2$ ranging
from 1.4 $nm$ to 411 $nm$.

From ellipsometry the ratio of p-polarized and s-polarized complex
Fresnel reflection coefficients is obtained \cite{Azz87,Tom99}:
\begin{equation}
\rho=\frac{r_p}{r_s}=\tan\Psi\;e^{i\Delta}, \label{Ellip}
\end{equation}
where $r_{p,s}$ are the corresponding reflection coefficients, and
the angles $\Psi$ and $\Delta$ are the raw data collected in a
measurement as functions of $\lambda$. All our films can be
considered as completely opaque, and can be described by the
reflection coefficients
\begin{equation}
r_p=\frac{<\varepsilon>\cos\vartheta-\sqrt{<\varepsilon>-\sin^2\vartheta}}
{<\varepsilon>\cos\vartheta+\sqrt{<\varepsilon>-\sin^2\vartheta}},\
\ \ r_s=\frac{\cos\vartheta-\sqrt{<\varepsilon>-\sin^2\vartheta}}
{\cos\vartheta+\sqrt{<\varepsilon>-\sin^2\vartheta}}, \label{refl}
\end{equation}
where $\vartheta$ is the angle of incidence and
$<\varepsilon>=<\varepsilon(\lambda)>$ is the "pseudo" dielectric
function of the films. The term "pseudo" is used here since the
films may not be completely isotropic or uniform; they are rough,
and may contain absorbed layers of different origin because they
have been exposed to air. If this is the case then the dielectric
function extracted from the raw data will be influenced by these
factors. For our films we expect high level of isotropy but they can
be not uniform in depth. It means that the dielectric function we
extract from the data will be averaged over the distance of the
order of the penetration depth. The roughness and absorbed layer can
have some significance in the visible and ultraviolet ranges but not
in the infrared, where the absorption on free electrons of metals is
very large. Moreover, the effect of roughness is expected to be
small since for all films $w$ is much smaller than the smallest
wavelength 137 $nm$. Because the infrared domain is the most
important for the Casimir force we will consider
$<\varepsilon(\lambda)>$ extracted from the raw data as a good
approximation for the dielectric function of the top layer of a gold
film.

The dielectric function is connected with the ellipsometric
parameter $\rho$ for an isotropic and uniform solid as
\begin{equation}
\varepsilon=\sin^2\vartheta\left[1+\tan^2
\vartheta\left(\frac{1-\rho}{1+\rho}\right)^2\right].
\label{Ell_eps}
\end{equation}
Instead of using $\varepsilon$ a material is often characterized by
the complex refractive index $\tilde{n}=n+ik=\sqrt{\varepsilon}$,
where $n$ is the refractive index and $k$ is the extinction
coefficient. Both descriptions are equivalent but the noise in the
data is weighted differently, and it can influence to some degree
the values of the Drude parameters (see next Section).

\begin{figure}[tbp]
\includegraphics[width=8.6cm]{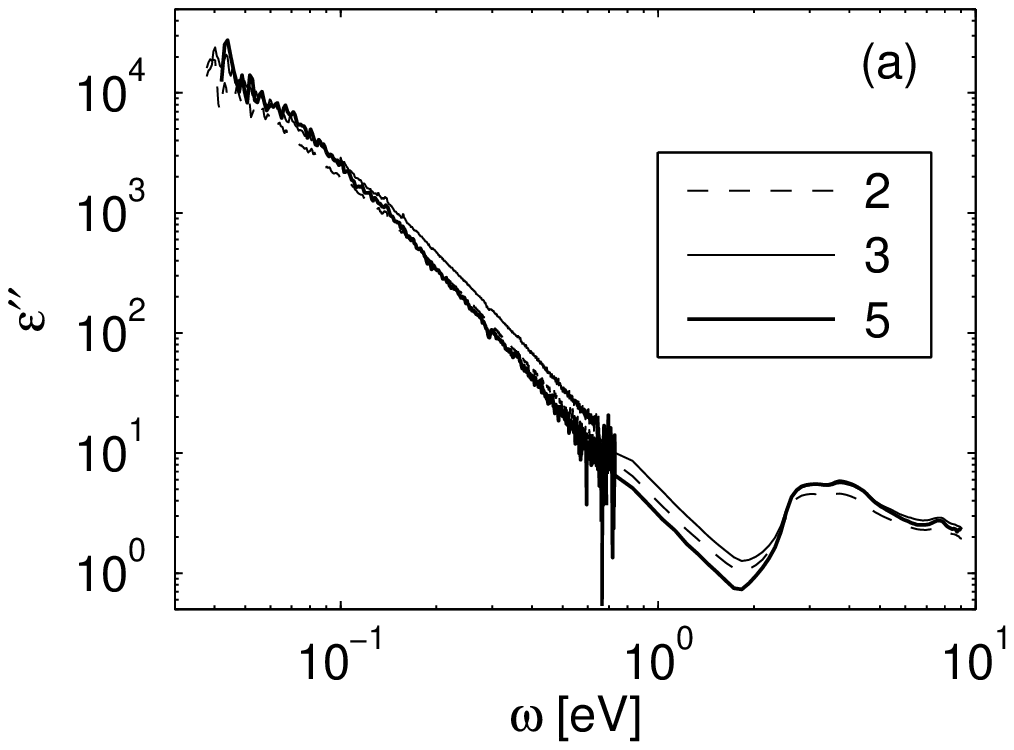}\newline
\includegraphics[width=8.6cm]{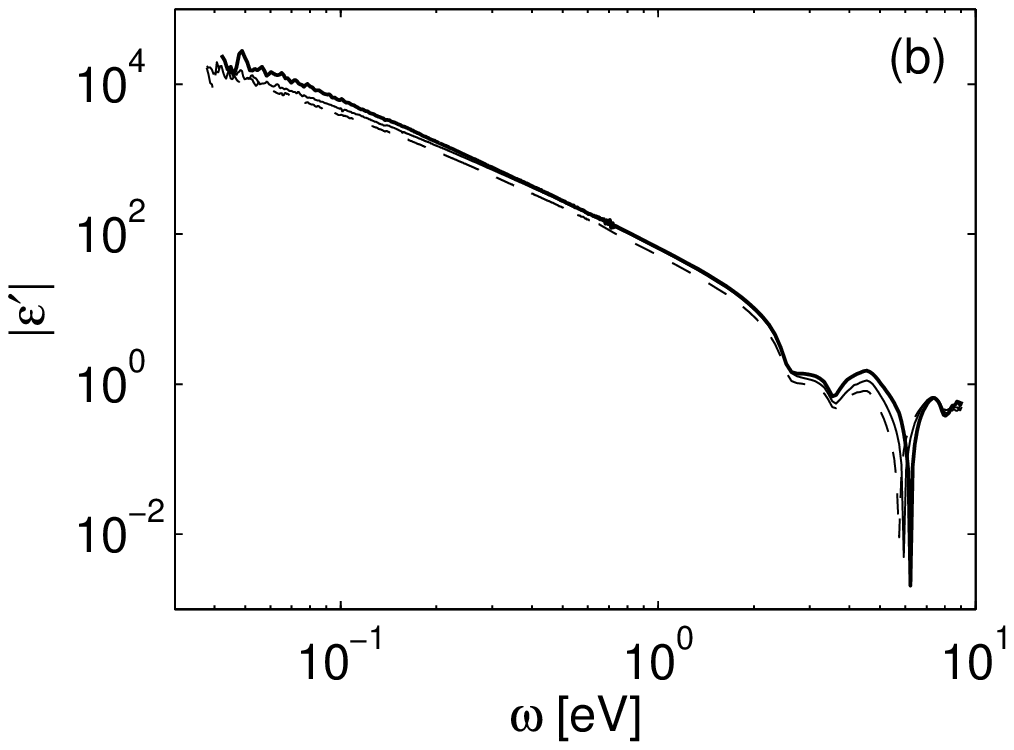}\newline
\caption{(a) Measured $\varepsilon^{\prime\prime}$ as a function of
frequency $\omega$. (b) The same for $|\varepsilon^{\prime}|$. }
\label{fig2}
\end{figure}

Figgure~\ref{fig2}(a) shows the experimental results for
$\varepsilon^{\prime\prime}(\omega)$ found via Eq.~(\ref{Ell_eps})
for 3 different films. One can see that the IR-VASE gives noisy
signal at both ends of the spectral interval. The noise is
significant for $\lambda>20\;\mu m$ but the number of points in this
range is not large, and the weight of these points for the
extraction of the Drude parameters or Kramers-Kronig analysis is
small. Around the interband transition (minimum of the the curves)
the smallest absorption is observed for the annealed sample on mica
indicating the smallest number of defects in the sample
\cite{Asp80}. On the contrary, this sample shows the largest
$|\varepsilon^{\prime}|$ in the infrared as one can see in
Fig.~\ref{fig1}(b). An important conclusion that can be drawn from
our measurements is the sample dependence of the dielectric
function. The log-log scale is not very convenient for having an
impression of this dependence. We present in Table \ref{tab1} the
values of $\varepsilon$ for all five samples at chosen wavelengths
$\lambda=1,5,10\;\mu m$. One can see that the real part of
$\varepsilon$ varies very significantly from sample to sample.

\begin{table}
\centering
\begin{tabular}{l||l|l|l}
Sample & $\lambda=1\;\mu m$ & $\lambda=5\;\mu m$ & $\lambda=10\;\mu m$ \\
\hline\hline 1, 400 $nm/Si$ & $-29.7+i 2.1$ & $-805.9+i 185.4$ & $-2605.1+i 1096.3$ \\
   2, 200 $nm/Si$ & $-31.9+i 2.3$ & $-855.9+i 195.8$ & $-2778.6+i 1212.0$ \\
   3, 100 $nm/Si$ & $-39.1+i 2.9$ & $-1025.2+i 264.8$ & $-3349.0+i 1574.8$ \\
   4, 120 $nm/Si$ & $-43.8+i 2.6$ & $-1166.9+i 213.9$ & $-3957.2+i 1500.1$ \\
   5, 120 $nm$/mica & $-40.7+i 1.7$ & $-1120.2+i 178.1$ & $-4085.4+i 1440.3$

\end{tabular}
\caption{Dielectric function for different samples at fixed
wavelengths $\lambda=1,5,10\;\mu m$.}\label{tab1}
\end{table}

\bigskip
The sample dependence of the dielectric function can be partly
attributed to different volume of voids in the films as was proposed
by Aspnes et al. \cite{Asp80}. To check this assumption we did
standard X-ray reflectivity (XRR) measurements
\cite{Sin88,Bra88,Als91} for the 100, 200, and 400 $nm$ $Au$ films
on $Si$, which were deposited at identical conditions. From this
kind of measurements one can draw information on the density of thin
films. For this purpose the Phillips Xpert diffractometer on the
$Cu\; K_{\alpha}$ radiation line $\lambda=1.54\;{\AA}$ was used. The
angle between the source and the surface was increased from 0.06 to
2 degrees. For hard X-rays the refractive index can be present as
$n\approx1-\delta$, where $\delta\ll 1$ and we neglect the imaginary
part of $n$ since absorption is small. For gold $\delta$ is given
\begin{equation}
\delta=\frac{Z}{2\pi}N_a\frac{e^2}{4\pi\varepsilon_0
mc^2}\lambda^2\approx
2.70\times10^{-6}\frac{N_e}{1\;cm^{-3}}\left(\frac{\lambda}{1\;\AA}\right)^2,
\label{theta_c}
\end{equation}
where $Z$ is the atomic number, $N_a$ is the number of atoms per
unit volume, $\varepsilon_0$ is the permittivity of vacuum, and
$N_e=ZN_a$ is the density of electrons.

\begin{figure}[tbp]
\includegraphics[width=8.6cm]{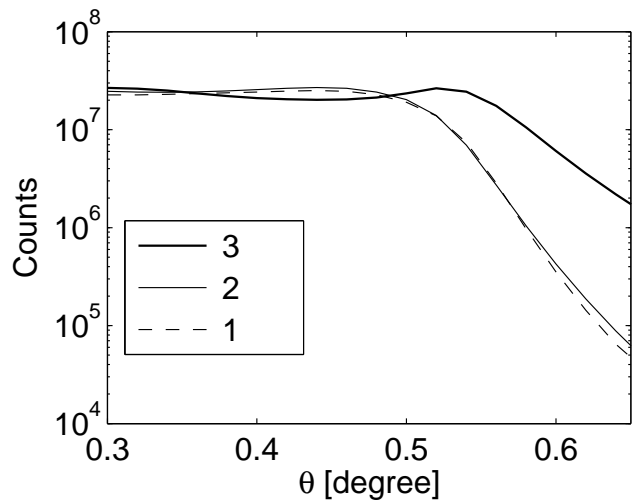}\newline
\caption{X-ray reflectivity (counts) vs the angle of incidence
(degrees). All 3 films were deposited at similar conditions.}
\label{fig3}
\end{figure}

Since X-rays refract away from the normal on the surface (refractive
index $n<1$), there exists a critical angle. Below this angle the
total reflection occurs. The critical angle $\theta_c$ can be
related to $\delta$ as $\theta_c\approx\sqrt{2\delta}$. Therefore,
by measuring $\theta_c$ one can provide knowledge of the electron
density $N_e$. The XRR results are shown in Fig.~\ref{fig3}. For
very small angles the reflectivity decreases (not shown here), which
may be due to the beam falling off the sample. Below the critical
angle the material reflectivity is generally not very well
understood, but it is of no concern to us since we are only
interested in the region around $\theta_c$. The transition region is
clearly visible on the graph. Above the critical angle the signal
drops very fast. The slope depends on the surface roughness. It is
clear from the graph that the 100 $nm$ film (sample 3) has the
largest critical angle and, therefore, the largest electron density
$N_e$. For bulk gold we have $N_e\approx 4.67\times
10^{24}\;cm^{-3}$. For our films we found from the critical angles
$N_e\approx (4.5\pm 0.8)\times 10^{24}\;cm^{-3}$ for sample 3 and
$N_e\approx (3.6\pm 0.8)\times 10^{24}\;cm^{-3}$ for the two other
films. The errors are rather large because $N_e\sim\theta_c^2$ and
the curve is not sharp at $\theta_c$. From this measurement we
cannot extract quantitative information, but qualitatively it agrees
with the suggestion of different volume of voids in the films.

\subsection{\label{Sec2B}Comparison with the existing data}
In the interband absorption region ($\omega>2.45\;eV$) there is
significant amount of data obtained by combined reflectance and
transmittance, ellipsometric spectroscopies on unannealed  or
annealed thin films or bulk samples measured in air or ultrahigh
vacuum. For comparison we have chosen the films by Th\`{e}ye
\cite{The70} evaporated in ultrahigh vacuum
($10^{-10}-10^{-11}\;Torr$) in the thickness range 10-25 $nm$ and
being well annealed. These films represent the bulk-like material.
The data were collected by measuring reflectance and transmittance
of the films. The Th\`{e}ye data became part of the handbook table
\cite{HB1} in the range $1<\omega<6\;eV$. The second choice is the
data by Johnson and Christy \cite{Joh72}. These films $25-50\;nm$
thick and they were thermally evaporated in vacuum  at $ 10^{-6}\;
Torr$. The data were collected for unannealed films from reflection
and transmission measurements. As the third choice we took the data
by Wang at al. \cite{Wan98}. The films were thermally evaporated at
pressure $10^{-5}\;Torr$. The data were collected with spectroscopic
ellipsometry for unannealed films of thickness 150 $nm$. The
preparation conditions and the method of measurement are similar to
that for our films.

\begin{figure}[tbp]
\includegraphics[width=8.6cm]{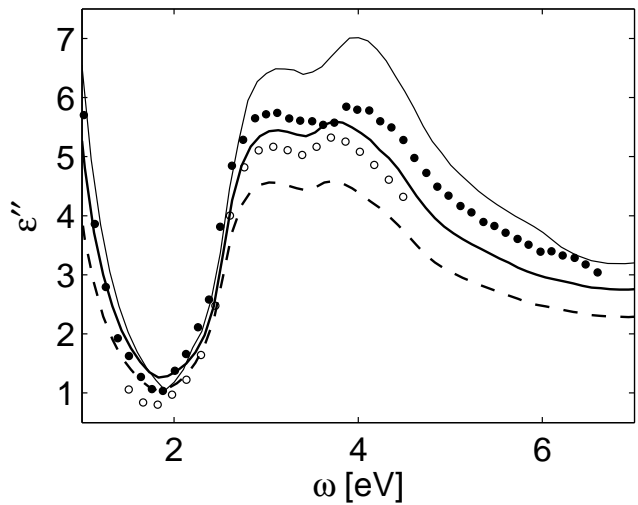}\newline
\caption{Imaginary part of the dielectric function in the interband
region obtained in different experiments. The thin solid line
represents the data by Th\`{e}ye \cite{The70}, the solid circles are
the data by Johnson and Christy \cite{Joh72}, and the open circles
are the data by Wang et al. \cite{Wan98}. The data of this study are
presented by thick solid (sample 3) and thick dashed (sample 2)
lines.} \label{fig_a}
\end{figure}

The imaginary part of the dielectric functions,
$\varepsilon^{\prime\prime}(\omega)$, for the chosen experiments are
shown in Fig.~\ref{fig_a} together with our data in the interband
range. One can see that in this range of frequencies our data are
rather typical. Only the Th\`{e}ye films, obtained in ultrahigh
vacuum and being well annealed, have significantly larger interband
absorption. It should be mentioned that the interband absorption
correlates with the film thickness: the thicker the film the less it
absorbs. The smallest absorption is observed for our 200 $nm$ film.
Wang et al. \cite{Wan98} deposited their films on bottom of Dove
prisms and measured the optical response on gold-air and gold-glass
interfaces. They have found that the interband absorption on the
gold-glass interface is larger than that on the gold-air interface.
In the former case the absorption is close to that observed by
Th\`{e}ye. It can be attributed to more dense parking of $Au$ atoms
nearby the substrate.

In the infrared and especially in mid- and far-infrared the
experimental data are sparse. In this range of frequencies both
$\varepsilon^{\prime}$ and $\varepsilon^{\prime\prime}$ were
measured only in a few studies \cite{Pad61,Mot64,Dol65}. Padalka and
Shklarevskii \cite{Pad61} thermally evaporated the films on glass at
pressure $10^{-5}\;Torr$. The films were not annealed; the thickness
of the films was not reported. They measured the optical constants
in the range $1<\lambda<12\;\mu m$. Motulevich and Shubin
\cite{Mot64} evaporated gold on polished glass at pressure $\sim
10^{-6}\;torr$. The investigated films were $0.5-1\;\mu m$ thick.
The samples were annealed in the same vacuum at $400^{\circ }\,C$.
The optical constants $n$ and $k$ were measured by the polarization
methods in the spectral range $1-12\;\mu m$. Dold and Mecke did not
describe the sample preparation carefully. It was only reported
\cite{Dol65} that the films were evaporated onto a polished glass
substrate, and measured in air by using an ellipsometric technique
in the range $1.25-14\;\mu m$. Presumably they were not annealed.
These data are included in the handbook \cite{HB1} table  in the
corresponding spectral range.

\begin{figure}[tbp]
\includegraphics[width=8.6cm]{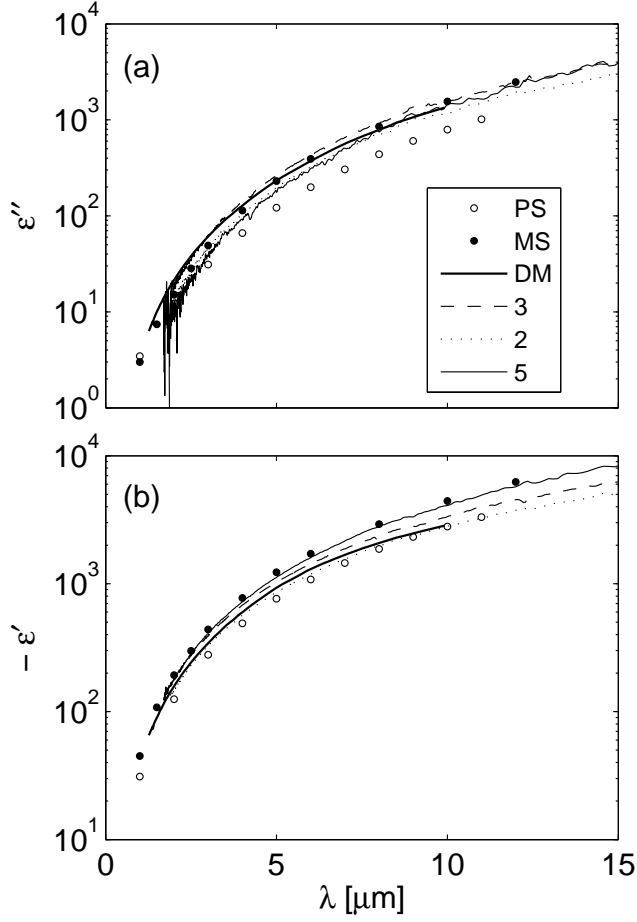}\newline
\caption{Comparison of existing low-frequency data for gold. Panel
(a) gives $\varepsilon^{\prime\prime}$ and panel (b) shows
$-\varepsilon^{\prime}$ as functions of the wavelength. The
literature data are marked as PM \cite{Pad61}, MS \cite{Mot64}, and
DM \cite{Dol65} (included in the handbook \cite{HB1}). Our data are
presented for samples 2, 3, and 5. } \label{fig_d}
\end{figure}

Fig.~\ref{fig_d} shows all the literature low-frequency data and
three of our films. The main conclusion that can be drawn from this
figure is that our films are typical in the sense of optical
properties. This is because all the films were deposited at similar
conditions. The annealed films by Motulevich and Shubin \cite{Mot64}
and our annealed sample 5 show the largest $-\varepsilon^{\prime}$.

\section{\label{Sec3}Analysis of the data}
The Casimir force given by the Lifshitz formula depends on the
dielectric function at imaginary frequencies: $\varepsilon(i\zeta)$.
This function cannot be measured directly in any experiment but with
the help of the Kramers-Kronig relation it can be expressed via the
observable function $\varepsilon^{\prime\prime}(\omega)$:
\begin{equation}
\varepsilon \left( i\zeta \right)=1+\frac{2}{\pi }\int\limits_{0}^{\infty }%
d\omega\frac{\omega \varepsilon ^{\prime \prime }\left( \omega
\right) }{\omega ^{2}+\zeta ^{2}}.  \label{K-K_imag}
\end{equation}
The experimental data available for
$\varepsilon^{\prime\prime}(\omega)$ are always restricted from low
and high frequencies. The low frequency cutoff $\omega_{cut}$ is
especially important in the case of metals. This is because for
metals $\varepsilon^{\prime\prime}$ is large at low frequencies
which contribute significantly to $\varepsilon(i\zeta)$
\cite{Sve00b}. Therefore, an important step in the evaluation of
$\varepsilon(i\zeta)$ is extrapolation of the dielectric function
$\varepsilon^{\prime\prime}(\omega)$ to low frequencies
$\omega<\omega_{cut}$, where the experimental data are not
accessible.

At low frequencies the dielectric function of metals can be
described by the Drude function
\begin{equation}
\varepsilon \left( \omega \right) =1-\frac{\omega _{p}^{2}}{\omega
\left( \omega +i\omega _{\tau }\right) },  \label{Drude}
\end{equation}
which is defined by two parameters, the plasma frequency $\omega
_{p}$ and the relaxation frequency $\omega _{\tau }$. Lambrecht and
Reynaud \cite{Lam00} fixed the plasma frequency using the relation
\begin{equation}
\omega _{p}^{2}=\frac{Ne^{2}}{\varepsilon _{0}m_{e}^{\ast }},
\label{Omp}
\end{equation}
where $N$ is the number of conduction electrons per unit volume, $e
$ is the charge and $m_{e}^{\ast }$ is the effective mass of
electron. The plasma frequency was evaluated assuming that each atom
gives one conduction electron and that the effective mass coincides
with the mass of free electron. The bulk density of $Au$ was used to
estimate $N$. The value of $\omega_p=9.0\;eV$ found in this way was
largely adopted by the community
\cite{Har00,Ede00,Cha01,Bres02,Dec03a,Dec03b,Dec05}. A relatively
close value of $\omega_p$ was found by Bennett and Bennett
\cite{Ben66} for carefully prepared films deposited in ultrahigh
vacuum \cite{Ben65}. However, it was stressed by these authors that
the reflectance for their films was always higher than the values
reported for films under standard vacuum conditions. The value
$\omega_p=9.0\;eV$ is close to the plasma frequency in a perfect
single crystal but the films used for measurement of the Casimir
force can contain defects, which are responsible for the reduction
of $\omega_p$. The most important defects are small voids, which
were observed in gold films with transmission electron microscopy
\cite{Llo77a,Llo77b}.

Special investigation of the influence of the defects was undertaken
by Aspnes et al. \cite{Asp80}, where it was stressed that films
grown at different conditions have considerably different
$\varepsilon^{\prime\prime}$ above the interband transition. The
spectra qualitatively differ only by scaling factors as one can see
in Fig.~\ref{fig_a}. The scaling behavior of the spectra was
attributed to different volumes of voids in the films prepared by
different methods \cite{Asp80}. Different kind of defects will also
contribute to the scattering of free electrons changing the
relaxation frequency $\omega_{\tau}$. Many researches stressed that
the conduction electrons are much more sensitive to slight changes
in the material structure \cite{Ben65,Asp80,The70,Kai02}. It is well
known, for example, that the resistivity of a film can be
significantly larger than the resistivity of the bulk material. At
any rate, so far we have to conclude that both Drude parameters must
be extracted from the optical data of the films, which are used for
the Casimir force measurement. Below we describe a few ways to
extract these parameters from our data.

\subsection{\label{Sec3A}Joint fit of $\varepsilon^{\prime}$ and $\varepsilon^{\prime\prime}$}
Separating the real and imaginary parts of the Drude function
(\ref{Drude}) one finds for $\varepsilon^{\prime}$ and
$\varepsilon^{\prime\prime}$:
\begin{equation}
\varepsilon ^{\prime }\left( \omega \right) =1-\frac{\omega
_{p}^{2}}{\omega ^{2}+\omega _{\tau }^{2}},\quad \varepsilon
^{\prime \prime }\left(\omega \right) =\frac{\omega _{p}^{2}\omega
_{\tau }}{\omega \left( \omega^{2}+\omega _{\tau }^{2}\right) }.
\label{DrudeRI}
\end{equation}
These equations have to be true below the interband transition
$\hbar\omega<2.45\;eV$ ($\lambda>0.5\;\mu m)$ \cite{The70}, but
because this transition is not sharp one has to do analysis at lower
frequencies. Practically Eqs.~(\ref{DrudeRI}) can be applied at
wavelengths $\lambda>2\;\mu m$ that coincides with the range of the
infrared ellipsometer. In this range Eqs.~(\ref{DrudeRI}) can be
compared with the optical data for both
$\varepsilon^{\prime}(\omega)$ and
$\varepsilon^{\prime\prime}(\omega)$. Minimizing deviations between
the data and theoretical expectations one can find the Drude
parameters $\omega_p$ and $\omega_{\tau}$. For example, for sample 3
($100\;nm,\ Au/Si$) we found $\omega_p=7.79\pm 0.01\;eV$ and
$\omega_{\tau}=48.8\pm 0.2\;meV$. Similar calculations for annealed
sample 5 ($120\;nm,\ Au/mica$) gave $\omega_p=8.37\pm 0.03\;eV$ and
$\omega_{\tau}=37.1\pm 0.5\;meV$. The statistical uncertainty of the
Drude parameters was found using a $\chi^2$ criterion for joint
estimation of two parameters \cite{Hag02}. For a given parameter the
error corresponds to the change $\Delta\chi^2 = 1$ when the other
parameter is kept constant. For all samples the values of the
parameters are collected in Table \ref{tab2}.

We found the parameters also in a slightly different way. One can
fit the complex refractive index
$\tilde{n}(\omega)=\sqrt{\varepsilon(\omega)}$ instead of the
dielectric function. In the Drude range, where nearly all absorption
is due to free electrons in the metal, $n(\omega)$ and $k(\omega)$
behave as
\begin{equation}
n(\omega)=\frac{\omega_p}{\sqrt{2}\omega_{\tau}}\frac{1}{\sqrt{1+y^2}}
\left[1+\frac{\sqrt{1+y^2}}{y}\right]^{-1/2}, \ \ \
k(\omega)=\frac{\omega_p}{\sqrt{2}\omega_{\tau}}\frac{1}{\sqrt{1+y^2}}
\left[1+\frac{\sqrt{1+y^2}}{y}\right]^{1/2},\label{Drude_n}
\end{equation}
where $y=\omega/\omega_{\tau}$. Then we can minimize deviations for
$n(\omega)$ and $k(\omega)$. The corresponding parameters for sample
3 are $\omega_p=7.94\pm 0.01\;eV$ and $\omega_{\tau}=52.0\pm
0.2\;meV$. For the annealed sample 5 they are $\omega_p=8.41\pm
0.02\;eV$ and $\omega_{\tau}=37.7\pm 0.4\;meV$. It can be noted that
within the statistical errors the parameters for the annealed film
are the same as those found by joint fit of $\varepsilon^{\prime}$
and $\varepsilon^{\prime\prime}$. However, for sample 3 this is not
the case.

\begin{figure}[tbp]
\includegraphics[width=8.6cm]{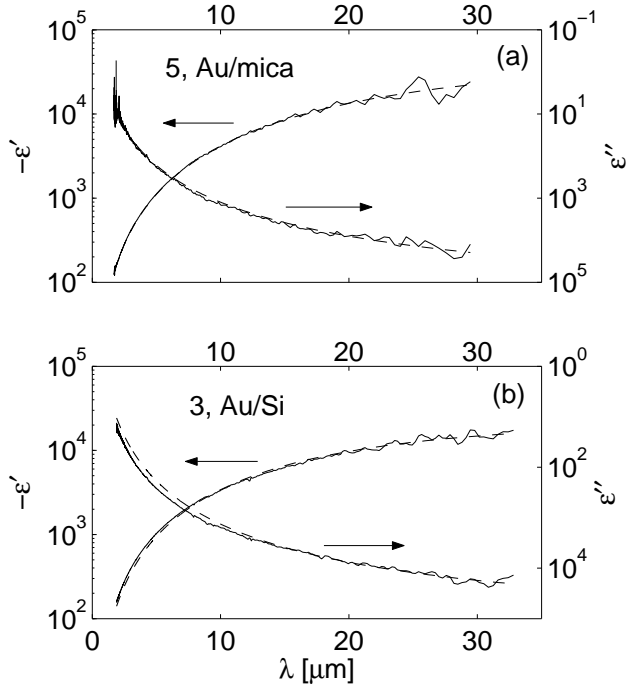}\newline
\caption{The infrared data as functions of the wavelength $\lambda$
for $\varepsilon^{\prime}$ and $\varepsilon^{\prime\prime}$ (solid
lines) and the best Drude fits (dashed lines) for two gold films.
Panel (a) shows the data for annealed sample 5 and panel (b) shows
the same for unannealed sample 3.} \label{fig_b}
\end{figure}

The Drude parameters should be the same in both cases but some
difference can appear due to the smaller contributing weight of low
frequencies when we perform minimization for $n$ and $k$ than that
from the minimization of $\varepsilon^{\prime}$ and
$\varepsilon^{\prime\prime}$. Figure~\ref{fig_b} shows the data for
$\varepsilon^{\prime}(\lambda)$ and
$\varepsilon^{\prime\prime}(\lambda)$ (solid lines) and the best
Drude fit (dashed lines) found by minimization of deviations for
$\varepsilon^{\prime}$ and $\varepsilon^{\prime\prime}$. Panel (a)
corresponds to sample 5 and panel (b) shows the data for sample 3.
The film on mica is described well by the Drude dielectric function,
but the film on Si demonstrates some deviations from the Drude
behavior at $\lambda<15\;\mu m$. It looks like an additional
absorption band. In principle, the anomalous skin effect can be
responsible for absorption in this range, but we found that this
effect can be observable only at smaller $\omega_{\tau}$. Additional
absorption in the Drude range is often observed  due to the tail of
the interband transition, but this is hardly the case because the
wavelength is too large.

\begin{figure}[tbp]
\includegraphics[width=8.6cm]{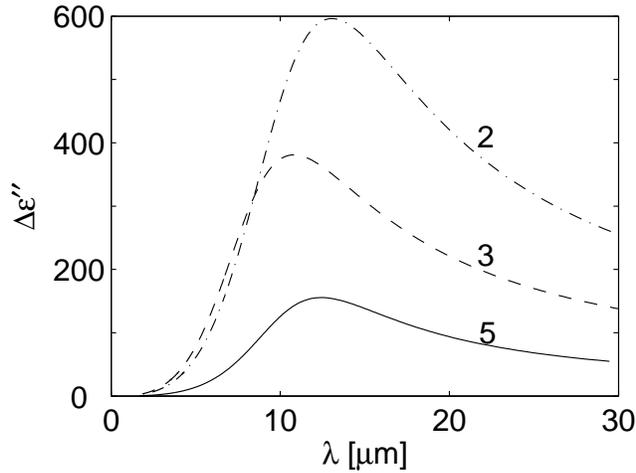}\newline
\caption{Additional absorption band for samples 2, 3, and 5 as a
function of the wavelength.} \label{fig_c}
\end{figure}

In absence of information on the origin of this absorption band we
did phenomenological analysis of our data with the dielectric
function, which includes an additional Lorentz oscillator:
\begin{equation}
\varepsilon \left( \omega \right)
=\varepsilon_{D}(\omega)+\frac{S\omega_0^2}{\omega_0^2-\omega^2-i\gamma\omega},
\label{new_band}
\end{equation}
where $\varepsilon_{D}$ is the Drude function (\ref{Drude}), $S$ is
the dimensionless oscillator strength, $\omega_0$ and $\gamma$ are
the central frequency and width of the band, respectively.
Fig.~\ref{fig_c} shows the difference
$\Delta\varepsilon^{\prime\prime}=
\varepsilon^{\prime\prime}-\varepsilon_D^{\prime\prime}$ as a
function of the wavelength $\lambda$. The smallest absorption is
realized for the annealed film on mica. It becomes larger for the
unannealed 100 $nm$ film on $Si$, and increases further for the 200
$nm$ film on $Si$. On the other hand, the central wavelength of the
band and its width do not change significantly from sample to sample
showing the common origin of the band for different samples. This
situation can be expected if the samples differ only by the density
of defects of the same kind. However, without additional
experimental information we cannot specify the exact nature of these
defects.

\subsection{Determination of the parameters using the Kramers-Kronig
relation \label{Sec3B}}

One of the disadvantages of the ellipsometric determination of the
dielectric function is that the method does not maintain
Kramers-Kronig (K-K) consistency. Therefore, it is important to
check the K-K relations for our data. To do the K-K analysis we also
have to extrapolate the dielectric function outside of the measured
frequency range. For metals extrapolation to low frequencies is
based again on the Drude dielectric function (\ref{Drude}). It means
that together with K-K consistency we will find the Drude parameters
for which this consistency is the best. This method for determining
the Drude parameters was used recently \cite{Pir06} for analysis of
the optical properties of gold samples, which are available in the
literature.

The K-K relation expresses $\varepsilon^{\prime}(\omega)$ as
integral over all frequencies of
$\varepsilon^{\prime\prime}(\omega)$ :
\begin{equation}\label{KKrel}
    \varepsilon^{\prime}(\omega)-1=\frac{2}{\pi
}P\int\limits_{0}^{\infty }dx\frac{x\varepsilon ^{\prime \prime
}\left( x\right) }{x^{2}-\omega ^{2}},
\end{equation}
where $P$ means the principal part of the integral. To use this
relation we have to define $\varepsilon^{\prime\prime}(\omega)$ at
all frequencies. Below the low-frequency cutoff we define it using
Eq.~(\ref{DrudeRI}). Above the high frequency cutoff (9 $eV$ for our
data) we enlarge the frequency range by using the handbook data
\cite{HB1} in the range $9<\omega<100\;eV$ and above 100~$eV$
$\varepsilon^{\prime\prime}(\omega)$ is extrapolated as
$A/\omega^3$. The constant $A$ is determined by matching
$\varepsilon^{\prime\prime}$ at $\omega=100\;eV$. The principal part
of the integral in (\ref{KKrel}) is calculated in the same way as in
\cite{Pir06}. $\varepsilon^{\prime}(\omega)$ was calculated in the
frequency range $0.01<\omega<100\;eV$ as a function of the Drude
parameters. This function was compared with the measured one and
minimization of deviations gave us the values of the parameters. The
points outside of the measured frequency range ($\omega<0.038\;eV$)
were compared with the prediction based on Eq.~(\ref{DrudeRI}).

\begin{figure}[tbp]
\includegraphics[width=8.6cm]{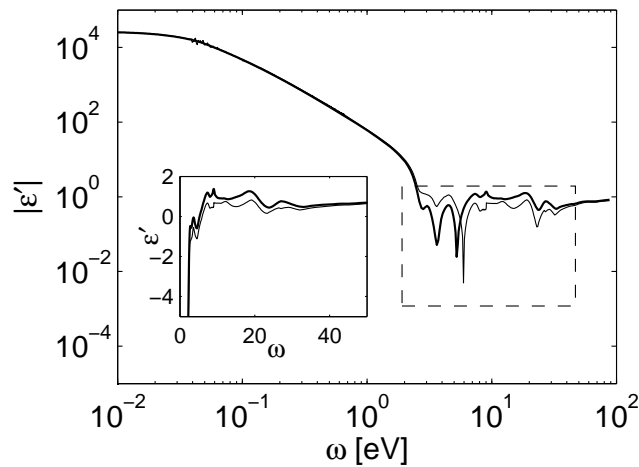}\newline
\caption{Dielectric function $\varepsilon^{\prime}$ as a function of
$\omega$ for sample 3. Thick line is the experimental curve
extrapolated to the frequencies lower than 0.038 $eV$ according to
the Drude model. Thin line is the same function calculated from the
K-K relation (\ref{KKrel}). Inset shows the behavior of
$\varepsilon^{\prime}$ inside of the dotted rectangular in the
linear scale.} \label{fig_e}
\end{figure}

For sample 3 it was found $\omega_p=7.80\;eV$ and
$\omega_{\tau}=47.9\;meV$. The experimental function
$\varepsilon^{\prime}(\omega)$ continued to lower frequency
according to the Drude model and its prediction based on the
relation (\ref{KKrel}) are shown in Fig.~\ref{fig_e} by thick and
thin lines, respectively. The agreement between the curves is rather
good. At high frequencies where $|\varepsilon^{\prime}|\sim 1$ the
logarithmic scale is not convenient for comparison. Instead we
present in the inset the region bounded by the dotted rectangular in
the linear scale. In this range the positions of the peaks are
reproduced very well, but their magnitude is slightly different.
This is because we used the data for $\varepsilon^{\prime\prime}$
above 9 $eV$ from the handbook, which did not match precisely to
those for our sample. Similar situation is realized for all the
other films. The Drude parameters for all samples are presented in
Table \ref{tab2}.

Alternatively one can use the experimental extinction coefficient
$k(\omega)$ to get the refractive index $n(\omega)$ using the K-K
relation between $n$ and $k$:
\begin{equation}\label{KK_refr}
    n(\omega)-1=\frac{2}{\pi
}P\int\limits_{0}^{\infty }dx\frac{xk\left( x\right) }{x^{2}-\omega
^{2}}.
\end{equation}
At frequencies for which the experimental data are not accessible we
define $k(\omega)={\rm Im}\sqrt{\varepsilon(\omega)}$, where
$\varepsilon(\omega)$ continued to low frequencies according to
(\ref{Drude}) and to high frequencies as
$\varepsilon(\omega)=1-\omega_p^2/\omega^2+iA/\omega^3$. The
constant $A$ again is chosen by matching
$\varepsilon^{\prime\prime}$ at the highest frequency
$\omega=100\;eV$.

In contrast with the K-K relation (\ref{KKrel}) now we cannot
present the contribution of low frequencies to the dispersion
integral (\ref{KK_refr}) in the analytic form. The Drude
representation (\ref{Drude_n}) for $k(\omega)$ at
$\omega\ll\omega_p$ is
\begin{equation}\label{kappa}
    k(\omega)=\frac{\omega_p}{\sqrt{2}\omega_{\tau}}\left[\frac{1}
    {y\sqrt{1+y^2}}+\frac{1}{1+y^2}\right]^{1/2},
\end{equation}
where $y=\omega/\omega_{\tau}$. This function of $y$ was
approximated by the forth order polynomial in $y$ in the range
$0<y<2$. With this polynomial the contribution of low frequencies to
(\ref{KK_refr}) was found analytically. Minimization of deviations
between the experimental values of $n(\omega)$ and the theoretical
predictions via (\ref{KK_refr}) gave the values of the Drude
parameters.

\begin{figure}[tbp]
\includegraphics[width=8.6cm]{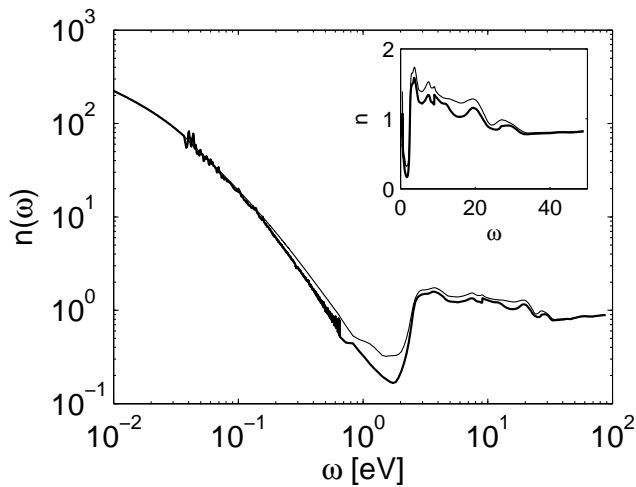}\newline
\caption{Refractive index $n$ as a function of $\omega$ for sample
3. Thick line is the experimental curve extrapolated to
$\omega<0.038\;eV$ according to the Drude model. Thin line is the
same function calculated from the K-K relation (\ref{KK_refr}). The
inset shows the behavior of $n$ in the linear scale in the range
$0.5<\omega<50\;eV$.} \label{fig_f}
\end{figure}

For sample 3 it was found $\omega_p=7.84\;eV$ and
$\omega_{\tau}=47.4\;meV$. The experimental and predicted curves for
$n(\omega)$ are shown in Fig.~\ref{fig_f} by thick and thin lines,
respectively. The inset shows the same functions in the linear plot
in the range $0.5<\omega<50\;eV$. Again we have a reasonable
agreement of the experiment and prediction on the basis of the K-K
relation.

\bigskip
The values of the Drude parameters found by different methods
described above are collected in Table \ref{tab2}. The statistical
errors in the parameters are rather small. They do not depend
significantly on the method and vary only slightly from sample to
sample. These errors are $0.01-0.03\; eV$ for $\omega_p$ and
$0.2-0.5\; meV$ for $\omega_{\tau}$. As one can see from Table
\ref{tab2} the values found by different methods do not agree with
each other within the statistical errors. This is because each
method treats noise in the data differently. We cannot give
preference to any specific method. Instead, we average the values of
the parameters determined by different methods, and define the
root-mean-square (rms) error of this averaging as uncertainty in the
parameter value. The averaged parameters and rms errors are given in
the last column of Table \ref{tab2}. Samples 1 and 2 have similar
Drude parameters, which cannot be resolved within the discussed
uncertainty, but all the other samples are clearly different.

\begin{table}
\centering
\begin{tabular}{l||l|l|l|l|l|l}
Sample & Parameter & Joint $\varepsilon^{\prime},\;\varepsilon^{\prime\prime}$ & Joint $n,\;k$ & K-K $\varepsilon^{\prime}$ & K-K $n$ & Average \\
\hline\hline 1 & $\omega_p\; [eV]$ & 6.70 & 6.87 & 6.88 & 6.83 & 6.82$\pm$0.08 \\
 400 $nm/Si$  & $\omega_{\tau}\; [meV]$ & 38.4 & 43.3 & 40.2 & 39.9 & 40.5$\pm$2.1\\
 \hline 2 & $\omega_p$ & 6.78 & 7.04 & 6.69 & 6.80 & 6.83$\pm$0.15 \\
 200 $nm/Si$ & $\omega_{\tau}$ & 40.7 & 45.3 & 36.1 & 36.0 & 39.5$\pm$4.4 \\
 \hline 3 & $\omega_p$ & 7.79 & 7.94 & 7.80 & 7.84 & 7.84$\pm$0.07 \\
 100 $nm/Si$ & $\omega_{\tau}$ & 48.8 & 52.0 & 47.9 & 47.4 & 49.0$\pm$2.1 \\
 \hline 4 & $\omega_p$ & 7.90 & 8.24 & 7.95 & 7.90 & 8.00$\pm$0.16 \\
 120 $nm/Si$ & $\omega_{\tau}$ & 37.1 & 41.4 & 35.2 & 29.2 & 35.7$\pm$5.1 \\
 \hline 5 & $\omega_p$ & 8.37 & 8.41 & 8.27 & 8.46 & 8.38$\pm$0.08 \\
 120 $nm/mica$ & $\omega_{\tau}$ & 37.1 & 37.7 & 34.5 & 39.1 &
 37.1$\pm$1.9
\end{tabular}
\caption{The Drude parameters determined by different methods
described in the text. In all cases the statistical errors in the
parameters are on the same level: $0.01-0.03\; eV$ for $\omega_p$
and $0.2-0.5\; meV$ for $\omega_{\tau}$. The last column shows the
values of the parameters averaged on different methods and the
corresponding rms errors.}\label{tab2}
\end{table}

\section{\label{Sec4} Sensitivity of the Casimir force to the optical properties of gold films}

In this Section we are going to discuss the Casimir force without
temperature or roughness corrections in order to concentrate on the
influence of the material optical properties on the force. In this
case the force between two similar parallel plates can be calculated
using the Lifshitz formula \cite{LP9}:
\begin{equation}
F_{pp}\left( a\right) =-\frac{\hbar}{2\pi^2
}\int\limits_{0}^{\infty}d\zeta\int\limits_{0}^{\infty
}dqqk_{0}\sum_{\mu=s,p}\frac{r_{\mu}^2e^{-2ak_0}}{1-r_{\mu}^2e^{-2ak_0}},
\label{Fpp}
\end{equation}
\noindent where ${\bf q}$ is the wave vector along the plates
($q=\left| {\bf q}\right| $). The formula includes the reflection
coefficients for two polarization states $\mu=s$ and $\mu=p$ which
are defined as
\begin{equation}
r_{s}=\frac{k_{0}-k_{1}}{k_{0}+k_{1}},\quad r_{p}=\frac{\varepsilon
\left( i\zeta \right) k_{0}-k_{1}}{\varepsilon \left( i\zeta \right)
k_{0}+k_{1}}  \label{rsp}
\end{equation}
\noindent with $k_{0}$ and $k_{1}$ being the normal components of
the wave vector in vacuum and metal, respectively:
\begin{equation}
k_{0}=\sqrt{\zeta^{2}/c^{2}+q^{2}},\quad k_{1}=\sqrt{\varepsilon
\left( i\zeta\right) \zeta^{2}/c^{2}+q^{2}}.  \label{k01}
\end{equation}

\subsection{\label{Sec4A}Dielectric function at imaginary frequencies}

To evaluate the force with the Lifshitz formula one has to know the
dielectric function of the material at imaginary frequencies
$\varepsilon(i\zeta)$, which is calculated via
$\varepsilon^{\prime\prime}(\omega)$ according to
Eq.~(\ref{K-K_imag}). Therefore, first we have to calculate
$\varepsilon(i\zeta)$ using our optical data. For this purpose let
us present $\varepsilon(i\zeta)$ as
\begin{equation}\label{eps2part}
    \varepsilon(i\zeta)=1+\varepsilon_{cut}(i\zeta)+\varepsilon_{exp}(i\zeta),
\end{equation}
where $\varepsilon_{cut}$ is calculated with the extrapolated
$\varepsilon^{\prime\prime}(\omega)$ in the unaccessible frequency
range $\omega<\omega_{cut}$, while $\varepsilon_{exp}$ is calculated
using the experimental data for
$\varepsilon^{\prime\prime}(\omega)$, according to the formulas:
\begin{equation}\label{part12}
    \varepsilon_{cut}(i\zeta)=\frac{2}{\pi}\int\limits_{0}^{\omega_{cut}}d\omega
    \frac{\omega\varepsilon^{\prime\prime}(\omega)}{\zeta^2+\omega^2},\
    \ \ \varepsilon_{exp}(i\zeta)=\frac{2}{\pi}\int\limits_{\omega_{cut}}^{\infty}d\omega
    \frac{\omega\varepsilon^{\prime\prime}(\omega)}{\zeta^2+\omega^2}.
\end{equation}
Strictly speaking, we included in $\varepsilon_{exp}$ the
extrapolation to high frequencies, $\omega>100\;eV$, but this is
justified because these frequencies do not play very significant
role. The high-frequency extrapolation of
$\varepsilon^{\prime\prime}(\omega)$ was done in the same way as in
Sec.~\ref{Sec3B} so as the calculation of the integral for
$\varepsilon_{exp}$. For the low-frequency extrapolation the Drude
model (\ref{DrudeRI}) was used. In this case the integral for
$\varepsilon_{cut}$ can be found analytically, and it yields
\begin{equation}\label{epscut}
    \varepsilon_{cut}(i\zeta)=\frac{2}{\pi}\frac{\omega_p^2}{\zeta^2-\omega_{\tau}^2}
    \left[\tan^{-1}\left(\frac{\omega_{cut}}{\omega_{\tau}}\right)-
    \frac{\omega_{\tau}}{\zeta}\tan^{-1}\left(\frac{\omega_{cut}}{\zeta}\right)\right].
\end{equation}
Note that there is no singularity here at $\zeta=\omega_{\tau}$.

\begin{figure}[tbp]
\includegraphics[width=8.6cm]{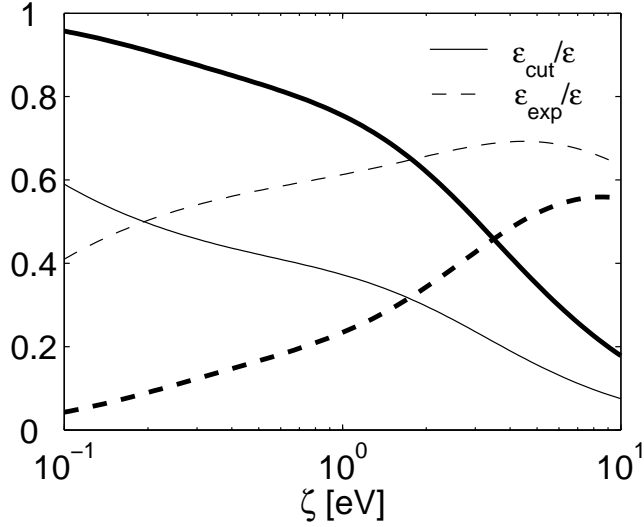}\newline
\caption{Relative contribution of low ($\omega<\omega_{cut}$) and
high ($\omega>\omega_{cut}$) frequencies to the dielectric function
$\varepsilon(i\zeta)$ for our sample 3 (thin lines) and for the
handbook data \cite{HB1} (thick lines). } \label{fig_10}
\end{figure}

It was already stressed that for metals $\varepsilon_{cut}$ gives an
important contribution to the dielectric function. Of course, it
depends on the value of $\omega_{cut}$. For all previous data this
value was around 0.125 $eV$. In this study $\omega_{cut}=0.038\;eV$
is about 3 times smaller, but still the contribution of
$\varepsilon_{cut}$ is significant. It can be seen from
Fig.~\ref{fig_10}, where the relative values
$\varepsilon_{cut}(i\zeta)/\varepsilon(i\zeta)$ and
$\varepsilon_{exp}(i\zeta)/\varepsilon(i\zeta)$ are presented for
sample 3 as thin lines. For comparison in the same plot we showed
(thick lines) $\varepsilon_{cut}$ and $\varepsilon_{exp}$ calculated
with the handbook data, and extrapolated to $\omega<0.125\;eV$ with
the Drude parameters $\omega_p=9.0\;eV$ and $\omega_{\tau}=35\;meV$.
It should be stressed that for our film the contribution from the
extrapolated region, $\varepsilon_{cut}$, dominates at
$\zeta<0.2\;eV$, while for the handbook data it dominates up to
$\zeta=4\;eV$. This is the result of reduced $\omega_{cut}$ for our
data. It means that the calculations based on our data are more
reliable because a smaller part of $\varepsilon(i\zeta)$ depends on
the extrapolation.

\begin{figure}[tbp]
\includegraphics[width=8.6cm]{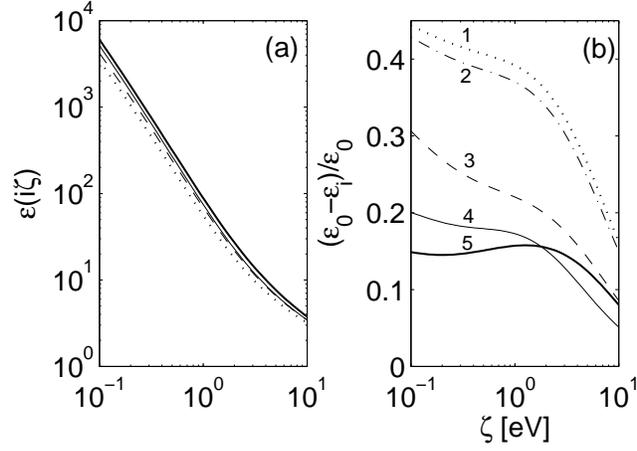}\newline
\caption{(a) The dielectric function as a function of the imaginary
frequency $\zeta$. The thin dotted, dashed, and solid lines
correspond to samples 2, 3, and 5, respectively. The thick line
gives $\varepsilon_0(i\zeta)$ that is typically used for the Casimir
force calculations. (b) Relative deviation of the dielectric
function of $i$-th sample, $\varepsilon_i$, from $\varepsilon_0$.}
\label{fig_11}
\end{figure}

The total dielectric functions $\varepsilon_i(i\zeta)$ ($i$ is the
number of the sample) are presented in Fig.~\ref{fig_11}(a) for
samples 2, 3, and 5. The results for samples 1 and 4 are not shown
for clarity. The thick solid line represents
$\varepsilon_0(i\zeta)$, which is typically used for the Casimir
force calculation. In this case the integral for $\varepsilon_{exp}$
is evaluated for $\varepsilon^{\prime\prime}(\omega)$ taken from the
handbook \cite{HB1}, where the cutoff frequency is
$\omega_{cut}=0.125\;eV$. The contribution of low frequencies,
$\varepsilon_{cut}$, is calculated with the Drude parameters
$\omega_p=9.0\;eV$ and $\omega_{\tau}=35\;meV$ \cite{Lam00}. In what
follows we are using $\varepsilon_0(i\zeta)$ as a reference case.

Important contribution to the Casimir force comes from the imaginary
frequencies around the characteristic frequency $\zeta_{ch}= c/2a$.
This frequency is in the range $0.1\lesssim\zeta_{ch}\lesssim
10\;eV$ when the distance is in the most interesting interval
$10\;nm \lesssim a\lesssim 1\;\mu m$. The frequency range in
Fig.~\ref{fig_11}(a), (b) was chosen accordingly. The logarithmic
scale in Fig.~\ref{fig_11}(a) does not give the feeling of actual
difference between the curves. The relative change in the dielectric
function
$\left[\varepsilon_0(i\zeta)-\varepsilon_i(i\zeta)\right]/\varepsilon_0(i\zeta)$
demonstrates much better the significance of actual optical
properties of the films. This change is shown in
Fig.~\ref{fig_11}(b) for all our films. The deviation of
$\varepsilon_i(i\zeta)$ from the imaginary material with the
dielectric function $\varepsilon_0(i\zeta)$, which is described by
the bulk Drude parameters and the handbook optical data, is
significant at all important frequencies and for all samples. Of
course, this deviation will be translated into the change in the
Casimir force. The curves in Fig.~\ref{fig_11}(b) were calculated
for the middle values of the Drude parameters in the last column of
Table~\ref{tab2}. The uncertainty of these parameters is responsible
for uncertainty of $\varepsilon_i(i\zeta)$. It is especially large
for samples 2 and 4. We will discuss it later in connection with the
uncertainty of the force.

\subsection{\label{Sec4B} The Casimir force}

To calculate the Casimir force the dielectric functions for all
samples, $\varepsilon_i(i\zeta)$, were found numerically in the
frequency range $0.01\;eV<\zeta<100\;eV$. At lower frequencies,
$\zeta<0.01\;eV$, they were extrapolated according to the Drude
model. At higher frequencies, $\zeta>100\;eV$, we extrapolated with
the function $\varepsilon_i(i\zeta)=1+A_i/\zeta^2$, where for each
sample the constant $A_i$ was chosen to match the value of
$\varepsilon_i(i\zeta)$ at $\zeta=100\;eV$.

\begin{figure}[tbp]
\includegraphics[width=8.6cm]{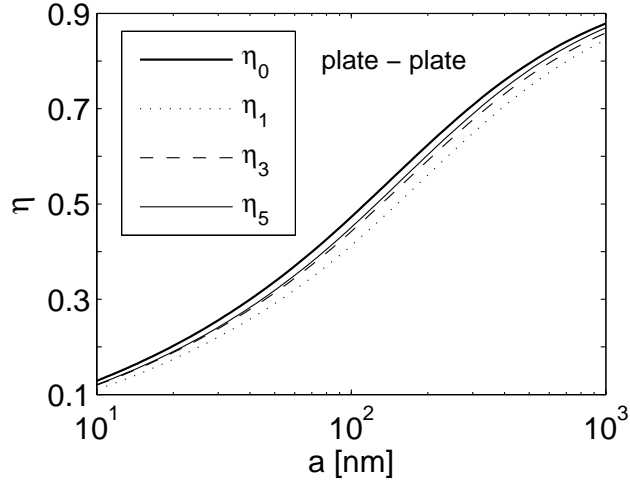}\newline
\caption{Reduction factor $\eta$ as a function of the separation $a$
for samples 1, 3, and 5. The thick line shows the reference result
calculated with $\varepsilon_{0}(i\zeta).$} \label{fig_12}
\end{figure}

\begin{figure}[tbp]
\includegraphics[width=8.6cm]{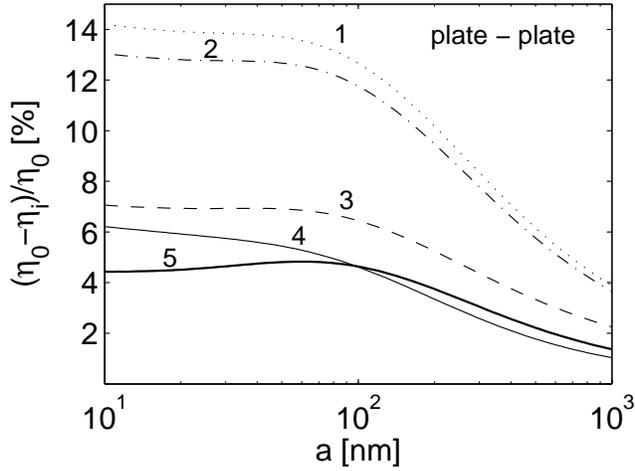}\newline
\caption{Relative deviation of the reduction factors for different
samples from the reference curve $\eta_0(a)$, which were evaluated
using the handbook optical data \cite{HB1} and the Drude parameters
$\omega_p=9\;eV$, $\omega_{\tau}=0.035\;eV$.} \label{fig_13}
\end{figure}

It is convenient to calculate not the force itself but so called the
reduction factor $\eta_{pp}$, which is defined as the ratio of the
force to the Casimir force between ideal metals:
\begin{equation}\label{eta_pp_def}
    \eta_{pp}(a)=\frac{F_{pp}(a)}{F^c_{pp}(a)},\ \ \ F^c_{pp}(a)=-\frac{\pi^2\hbar c}{240
    a^4}.
\end{equation}
For convenience of the numerical procedure one can make  an
appropriate change of variables in Eq.~(\ref{Fpp}) so that the
reduction factor can be presented in the form
\begin{equation}\label{eta_pp}
    \eta_{pp}(a)=\frac{15}{2\pi^4}\sum\limits_{\mu=s,p}\int\limits_{0}^{1}dx
    \int\limits_{0}^{\infty}\frac{dyy^3}{r^{-2}_{\mu}e^y-1},
\end{equation}
where the reflection coefficients as functions of $x$ and $y$ are
defined as
\begin{equation}\label{refl_xy}
    r_s=\frac{1-s}{1+s},\ \ \
    r_p=\frac{\varepsilon(i\zeta_{ch}xy)-s}{\varepsilon(i\zeta_{ch}xy)+s},
\end{equation}
with
\begin{equation}\label{s_def}
    s=\sqrt{1+x^2\left[\varepsilon(i\zeta_{ch}xy)-1\right]},\ \ \ \zeta_{ch}=\frac{c}{2a}.
\end{equation}
The integral (\ref{eta_pp}) was calculated numerically with
different dielectric functions $\varepsilon_i(i\zeta)$ with a
precision of $10^{-6}$. The results are presented in
Fig.~\ref{fig_12} for samples 1, 3, and 5.  The reference curve
(thick line) calculated with $\varepsilon_0(i\zeta)$ is also shown
for comparison. It represents the reduction factor, which is
typically used in the precise calculations of the Casimir force
between gold surfaces. One can see that there is significant
difference between this reference curve and those that correspond to
actual gold films. To see the magnitude of the deviations from the
reference curve, we plot in Fig.~\ref{fig_13} the ratio
$(\eta_0-\eta_i)/\eta_0$ as a function of distance $a$ for all five
samples.

At small  distances the deviations are more sensitive to the value
of $\omega_p$. At large distances the sample dependence becomes
weaker and more sensitive to the value of $\omega_{\tau}$. For
samples 1 and 2, which correspond to the 400 $nm$ and 200 $nm$ films
deposited on the Si substrate, the deviations are especially large.
They are 12-14\% at $a<100\;nm$ and stay considerable even for the
distances as large as 1 $\mu m$. Samples 3, 4, and 5 have smaller
deviations from the reference case but even for these samples the
deviations are as large as 5-7\%.

\begin{figure}[tbp]
\includegraphics[width=8.6cm]{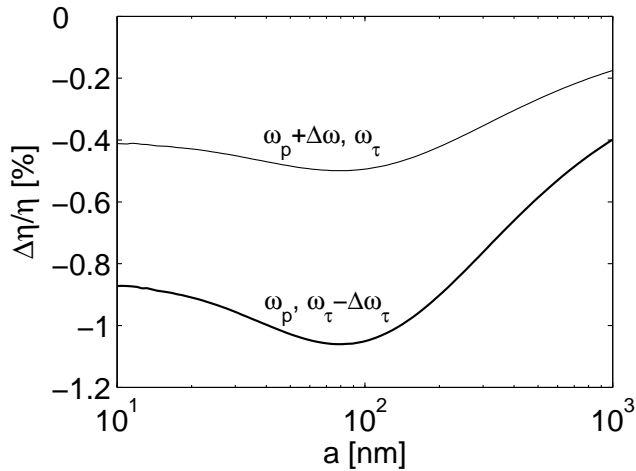}\newline
\caption{Relative variation of the reduction factor (see definition
in the text) for the Drude parameters at the sides of the error
intervals. The results are presented for sample 4.} \label{fig_14}
\end{figure}

We calculated also how uncertainty in the Drude parameters
influences the uncertainty in the force. For this purpose we
calculated the reduction factor $\eta$ at the borders of the error
intervals: $[\omega_p+\Delta\omega_p,\; \omega_{\tau}]$ and
$[\omega_p,\; \omega_{\tau}-\Delta\omega_{\tau}]$, where
$\Delta\omega_p$ and $\Delta\omega_{\tau}$ are shown as the errors
in the last column of Table~\ref{tab2}. The results were compared
with $\eta$ calculated with the middle values of the parameters
$[\omega_p,\;\omega_{\tau}]$. The maximal deviations were found for
sample 4. The relative deviations for this sample,
$\Delta\eta/\eta$, are shown in Fig.~\ref{fig_14} as functions of
the separation $a$. These deviations, $\Delta\eta/\eta$, are defined
as
\begin{equation}\label{del_eta}
    \frac{\Delta\eta}{\eta}=\frac{\eta(\omega_p,\omega_{\tau},a)-
    \eta(\omega_p+\delta\omega_p,\omega_{\tau}+\delta\omega_{\tau},a)}
    {\eta(\omega_p,\omega_{\tau},a)},
\end{equation}
where the variations of the plasma and relaxation frequencies,
$\delta\omega_p$ and $\delta\omega_{\tau}$, give the maximal effect
on the reduction factor when $\delta\omega_p=\Delta\omega_p$ and
$\delta\omega_{\tau}=-\Delta\omega_{\tau}$.

In most of the experiments the force was measured between a gold
covered sphere and a plate. In the case of the sphere-plate
interaction we define the reduction factor as
\begin{equation}\label{eta_sp_def}
    \eta_{sp}(a)=\frac{F_{sp}(a)}{F^c_{sp}(a)},\ \ \ F^c_{sp}(a)=-\frac{\pi^3R\hbar c}{360
    a^3},
\end{equation}
where $R$ is the sphere radius. The expression convenient for
numerical calculations in this case is the following:
\begin{equation}\label{eta_sp}
    \eta_{sp}(a)=-\frac{45}{2\pi^4}\sum\limits_{\mu=s,p}\int\limits_{0}^{1}dx
    \int\limits_{0}^{\infty}dyy^2\ln\left(1-r^2_{\mu}e^{-y}\right),
\end{equation}
where the reflection coefficients are defined in (\ref{refl_xy}).
Although in the case of sphere and plate the reduction factors are
different from those between two plates, the qualitative behavior
with the separation is the same. The relative deviations
$(\eta_0-\eta_i)/\eta_0$ for the sphere and plate are very close to
those shown in Fig.~\ref{fig_13}, and we do not show this additional
plot.

\section{Conclusions \label{Sec5}}

In this paper we analyzed the optical properties of gold films
deposited with different techniques on silicon or mica substrates.
The optical responses were measured ellipsometrically in a wide
range of wavelengths. In the mid-infrared our data cover the range
of wavelengths up to 33~$\mu m$, which is larger than in all
previous studies ($\sim 10~\mu m$), where both the real and
imaginary parts of the dielectric function were measured. The data
unambiguously demonstrate the sample dependence of the dielectric
function. It was found that thicker films are inferior in the
optical sense than thin (but opaque) films. It is probably connected
with more dense parking of atoms nearby the substrate. The X-ray
reflectivity measurements qualitatively support this interpretation.
We also observed some difference between samples deposited in the
electron-beam evaporator or in the thermal evaporator, but it is
possible that this difference is connected with different sublayers
(titanium or chromium, respectively) on $Si$ substrates.

The Casimir force depends on the dielectric function at imaginary
frequencies, $\varepsilon(i\zeta)$, which is expressed via the
measurable function $\varepsilon^{\prime\prime}(\omega)$ by the
dispersion relation (\ref{K-K_imag}). For metals low frequencies
play a very important role in this relation. However, in experiments
there is a low frequency cutoff, $\omega_{cut}$, so that the data at
smaller frequencies are not accessible. To make precise evaluation
of the force, one has to extrapolate the measured
$\varepsilon^{\prime\prime}(\omega)$ to the frequencies
$\omega<\omega_{cut}$. For noble metals like gold it can be done
with the help of the Drude model. The parameters of the Drude model
$\omega_p$ and $\omega_{\tau}$ have to be extracted from the optical
data at $\omega>\omega_{cut}$. In this paper we found the Drude
parameters for our films using joint fit of
$\varepsilon^{\prime}(\omega)$ and
$\varepsilon^{\prime\prime}(\omega)$ and independently using the
Kramers-Kronig consistency of the data. As an alternative we did the
same analysis for the refraction index and the extinction
coefficient. The results collected in Table~\ref{tab2} exhibiting
significant variation of the parameters from sample to sample.
Moreover, the parameters are slightly different when different
methods to fit the data are used. This is because the noise in the
data are weighted differently for different methods of parameter
determination. We defined the values and the errors in the
parameters as the averaged values over different methods and the
corresponding rms errors, respectively.

The contribution of the extrapolated part of
$\varepsilon^{\prime\prime}(\omega)$ to $\varepsilon(i\zeta)$ is
considerably smaller for our data than for the handbook data as one
can see in Fig.~\ref{fig_10}. The reason is that in our case the
cutoff frequency $\omega_{cut}=0.0378\;eV$ is smaller than that for
the handbook data, where $\omega_{cut}=0.125\;eV$. Nevertheless,
this contribution is still significant and has to be carefully
considered. As a reference curve for $\varepsilon(i\zeta)$ we have
chosen $\varepsilon_0(i\zeta)$, which was calculated with the
handbook data for $\omega>0.125\;eV$, and extrapolated for smaller
frequencies with the Drude parameters $\omega_p=9.0\;eV$ and
$\omega_{\tau}=35\;meV$. This curve is typically used for the
calculation of the Casimir force. For our films we have found that
$\varepsilon_i(i\zeta)$ was always smaller than the reference curve,
and the relative deviation is larger than 15\% at $\zeta\sim 1\;eV$
(see Fig.~\ref{fig_11}(b)).

Indeed, the Casimir force evaluated for our films is considerably
smaller from that calculated with the reference curve
$\varepsilon_0(i\zeta)$. The smallest deviation is realized for
sample 5 (120 $nm\ Au/mica$) and estimated to be 4-5\% in the
distance range $a\lesssim 200\;nm$. For thicker films (sample 1 or
2) it can be as large as 14\% (see Fig.~\ref{fig_13}). At large
separations the sample dependence becomes weaker, but it is above
1\% even at $a=1\;\mu m$.

Our main conclusion is that actual optical properties of the
materials used for the measurement of the Casimir force are very
important for comparison between theory and experiment. These
properties have to be measured for the same materials, and not taken
from a handbook for a material of the same chemical nature but
possibly of very different microstructure. It is concluded that
prediction of the Casimir force  with a precision better than 10\%
must be based on the material optical response measured from visible
to mid-infrared range.


\begin{references}
\bibitem{Cas48}  H. B. G. Casimir, Proc. K. Ned. Akad. Wet. {\bf 51},
793 (1948).

\bibitem{Lam97}  S. K. Lamoreaux, Phys. Rev. Lett. {\bf 78}, 5 (1997);
{\bf 81}, 5475 (1998).

\bibitem{Moh98}  U. Mohideen and A. Roy, Phys. Rev. Lett. {\bf 81},
4549 (1998); A. Roy, C.-Y. Lin, and U. Mohideen, Phys. Rev. D {\bf
60}, 111101(R) (1999).

\bibitem{Har00}  B. W. Harris, F. Chen, and U. Mohideen, Phys. Rev. A
{\bf 62}, 052109 (2000).

\bibitem{Cha01}  H. B. Chan, V. A. Aksyuk, R. N. Kleiman, D. J.
Bishop, and F. Capasso, Science {\bf 291}, 1941 (2001); Phys. Rev.
Lett. {\bf 87}, 211801 (2001).

\bibitem{Dec03a}  R. S. Decca, D. L\'{o}pez, E. Fischbach, and D. E.
Krause, Phys. Rev. Lett. {\bf 91}, 050402 (2003).

\bibitem{Dec03b}  R. S. Decca, E. Fischbach, G. L. Klimchitskaya, D. E.
Krause, D. L\'{o}pez, and V. M. Mostepanenko, Phys. Rev. D {\bf 68},
116003 (2003).

\bibitem{Dec05} R. S. Decca , D. L\'{o}pez, E. Fischbach, G. L. Klimchitskaya,
 D. E. Krause, and V. M. Mostepanenko, Ann. Phys. \textbf{318} 37 (2005).

\bibitem{Bres02}  G. Bressi, G. Carugno, R. Onofrio, and G. Ruoso,
Phys. Rev. Lett. {\bf 88}, 041804 (2002).

\bibitem{Ede00}  T. Ederth, Phys. Rev. A 62, 062104 (2000).

\bibitem{Lif56}  E. M. Lifshitz, Zh. Eksp. Teor. Fiz. {\bf 29}, 94 (1956)
[Sov. Phys. JETP {\bf 2}, 73 (1956)].

\bibitem{DLP} I.E.~Dzyaloshinskii, E.M.~Lifshitz and L.P.~Pitaevskii,
Advances in Physics {\bf 38}, 165 (1961).

\bibitem{LP9}  E. M. Lifshitz and L. P. Pitaevskii, {\it Statistical
Physics, Part 2} (Pergamon Press, Oxford, 1980).

\bibitem{Kli96}  G. L. Klimchitskaya and Yu. V. Pavlov, Int. J. Mod.
Phys. A {\bf 11}, 3723 (1996).

\bibitem{Gen03} C. Genet, A. Lambrecht, P. Maia Neto and S. Reynaud,
Europhys. Lett. {\bf 62}, 484 (2003).

\bibitem{Mia05} P. A. Maia Neto, A. Lambrecht and S. Reynaud, Phys. Rev A {\bf 72}, 012115
(2005).

\bibitem{Kli00}  G. L. Klimchitskaya, U. Mohideen, and V. M.
Mostepanenko, Phys. Rev. A {\bf 61}, 062107 (2000).

\bibitem{HB1}  {\it Handbook of Optical Constants of Solids}, edited by
E. D. Palik (Academic Press, 1995).

\bibitem{Wea81}  J. H. Weaver, C. Krafka, D. W. Lynch, and E. E.
Koch, {\it Optical Properties of Metals, Part II, Physics Data No.
18-2} (Fachinformationszentrum Energie, Physik, Mathematik,
Karsruhe, 1981).

\bibitem{Lam99b} S. K. Lamoreaux, Phys. Rev. A {\bf 59}, R3149
(1999).

\bibitem{Lam99a} S. K. Lamoreaux, Phys. Rev. Lett. {\bf 83}, 3340
(1999).

\bibitem{Sve00a}  V. B. Svetovoy and M. V. Lokhanin, Mod. Phys. Lett.
A {\bf 15}, 1013 (2000).

\bibitem{Sve00b}  V. B. Svetovoy and M. V. Lokhanin, Mod. Phys. Lett.
A {\bf 15}, 1437 (2000).

\bibitem{Pir06} I. Pirozhenko, A. Lambrecht, and V.B. Svetovoy, New
J. Phys. {bf 8} 238 (2006).

\bibitem{Mea94} P. Meakin Phys. Rep. {\bf 235}, 1991 (1994).

\bibitem{Kri95} J. Krim and G. Palasantzas, Int. J. of Mod. Phys. B {\bf9}, 599
(1995).

\bibitem{Zho01} Y. -P. Zhao, G. -C. Wang, and T. -M. Lu, { \it Characterization of amorphous
and crystalline rough surfaces-principles and applications},
(Experimental Methods in the Physical Science Vol. 37, Academic
Press, 2001).

\bibitem{Zwo07} P. van Zwol, G. Palasantzas, and J. Th. M. De
Hosson, Appl. Phys. Lett. {\bf 91}, 144108 (2007).

\bibitem{Woo} http://www.JAWoollam.com

\bibitem{Azz87} R. M. A. Azzam and N. M. Bashara, {\it Ellipsometry and Polarized Light} (Amsterdam, North Holland, 1987).

\bibitem{Tom99} H. G. Tompkins and W. A. McGahan, {\it Spectroscopic Ellipsometry and Reflectometry} (New York, Wiley, 1999).

\bibitem{Asp80}  D. E. Aspnes, E. Kinsbron, and D. D. Bacon, Phys.
Rev. B {\bf 21}, 3290 (1980).

\bibitem{Sin88} S. K. Sinha, E. B. Sirota, S. Garoff, and H. B. Stanley, Phys. Rev. B {\bf 38} 2297
(1988).

\bibitem{Bra88} A. Braslau, P. S. Pershan, G. Swislow, B. M. Ocko, and J. Als-Nielsen, Phys. Rev. A {\bf 38} 2457 (1988).

\bibitem{Als91} J. Als-Nielsen, {\it Handbook of synchrotron radiation} (North Holland New York,
1991).

\bibitem{The70}  M.-L. Th\`{e}ye, Phys. Rev. B {\bf 2}, 3060 (1970).

\bibitem{Joh72}  P. B. Johnson and R. W. Christy, Phys. Rev. B {\bf 6},
4370 (1972).

\bibitem{Wan98}  Yu Wang et al., Thin Solid Films {\bf 313}, 232
(1998).

\bibitem{Dol65}  B. Dold and R. Mecke, Optik, {\bf 22}, 435 (1965).

\bibitem{Mot64}  G. P. Motulevich and A. A. Shubin, Zh. Eksp. Teor.
Fiz. {\bf 47}, 840 (1964) [Soviet Phys. JETP {\bf 20}, 560 (1965)].

\bibitem{Pad61}  V. G. Padalka and I. N. Shklyarevskii, Opt.
Spektroskopiya {\bf 11}, 527 (1961) [Opt. Spectry. (USSR) {\bf 11},
285 (1961)].

\bibitem{Lam00}  A. Lambrecht and S. Reynaud, Eur. Phys. J. D {\bf 8},
309 (2000).

\bibitem{Ben66}  H. E. Bennett and J. M. Bennett, in \emph{Optical
Properties and Electronic Structure of Metals and Alloys}, edited by
F. Ab\'{e}l\`{e}s (North-Holland Publ., Amsterdam, 1966).

\bibitem{Ben65}  J. M. Bennett and E. J. Ashley, Appl. Opt. {\bf 4}, 221
(1965).

\bibitem{Llo77a} J. A. Lloyd and S. Nakahara, J. Appl. Phys. {\bf 48},
5092 (1977).

\bibitem{Llo77b} J. A. Lloyd and S. Nakahara, J. Vac. Sci. Technol. {\bf 14},
655 (1977).

\bibitem{Kai02} N. Kaiser, Appl. Opt. {\bf 41}, 3053 (2002).

\bibitem{Hag02} K. Hagiwara et al., Phys. Rev. D {\bf 66}, 010001 (2002) .


\end{references}
\end{document}